\documentclass[]{spie}
\usepackage[]{graphicx}

\title{Compact Stabilized Semiconductor Laser for  Frequency Metrology}

\author{
 W. Liang, V. S. Ilchenko, D. Eliyahu, E. Dale, A. A. Savchenkov, D. Seidel, A. B. Matsko, and L. Maleki
\skiplinehalf OEwaves Inc., 465 N. Halstead Street, Suite 140, Pasadena, California, 91107, USA}

\authorinfo{Send correspondence to A.B.Matsko:
Andrey.Matsko@oewaves.com}

\begin{document}

\maketitle

\begin{abstract}
We report on the development of a frequency modulatable 795~nm semiconductor laser based on self-injection locking to a high quality factor whispering gallery mode microresonator. The laser is characterized with residual amplitude modulation below -80~dB and frequency noise better than 300~Hz/Hz$^{1/2}$ at offset frequencies ranging from 100~Hz to 10~MHz.  The frequency modulation (FM) speed and span of the laser exceed 1~MHz and 4~GHz, respectively. Locking of the laser to Doppler-free saturated absorption resonance of $^{87}$Rb D$_1$ line is demonstrated and frequency stability below $10^{-12}$ is measured for integration time spanning from 1~s to 1~day. The architecture demonstrated in this study is suitable for realization of frequency modulatable lasers at any wavelength.
\end{abstract}


\section{Introduction}

Laser wavelength modulation (WM) \cite{moses77ol} and frequency modulation (FM) \cite{bjorklund80ol,hall81apl} spectroscopy is a versatile technique to selectively detect variations of optical absorption caused by atomic and molecular transitions. This approach is widely utilized in a variety of applications, including FM spectroscopy, and atomic clocks. In these and related applications the signal contrast is an important parameter since it determines the ultimate signal to noise ratio, which is a key factor in the achievable performance.  The sensitivity limit of WM/FM spectroscopy is  also controlled by the background signals originating from residual amplitude modulation (RAM) \cite{whittaker85josab,gehrtz85josab,wong85josab,moerner89prl,ishibashi02qels,duburck03tim,duburck04mst,duburck05el,jaatinen09mst,sathian12ao} as well as the inherent noise of the laser, which sets the frequency resolution limit.

The detection of low contrast absorption features in FM spectroscopy is aided by fast modulation of frequency at rates comparable with the bandwidth of the resolved features \cite{supplee94ao}. The low contrast spectroscopic signal is extracted by self-homodyning lock-in methods, which improve the signal-to-noise ratio by essentially subtracting the contribution of base-band amplitude and frequency noise of the laser. Resolution improvements of 40~dB and more are easily obtained by combining high-performance stabilized lasers with external frequency or phase (acousto-optic and electro-optic) modulators.  Externally modulated FM spectroscopy systems can resolve parts per million sub-Doppler and multiphoton absorption features,  reaching sub-Hz resolution to achieve the high level of signal-to-noise ratio that is important in optical frequency metrology, pectroscopy, and optical clocks.

Semiconductor lasers are attractive for these applications \cite{hinkley71s,bjorklund81pra,pokrowsky83oc,carlisle89ao,zhu97josab}, especially where compact spectroscopy systems are desired. Diode lasers are available at many optical wavelengths and are directly tunable/ditherable via either modulation of their injection current or thier temperature.   Resolution, wavelength precision, and sensitivity of diode laser-based FM spectroscopy is limited by their relatively broad linewidth ($\approx$ 1~MHz), high frequency noise, as well as the significant RAM of the laser. This occurs in spite of  high rates and spans of modulation that are achievable with direct modulation of Fabry-Perot (FP) and distributed feedback (DFB) lasers. The linewidth can be improved and frequency modulation spans can by increased and RAM can be reduced by using external cavity frequency stabilization technique. The optimized laser linewidth achieved with this approach is on the order of tens of kHz in the common external cavity diode lasers used in Littman or Littrow configurations \cite{littman78ao,liu81ol,harvey91ol,lecomte00ao,ricci95oc,hawtorn01rsi}.  It is uncommon to achieve a combination of a narrow linewidth (1-10~kHz), broad dithering span (exceeding 1~GHz), and high dithering frequency (10~kHz and higher rates) desirable for many spectroscopic applications.

In this work we report on demonstration of a semiconductor laser with substantially better characteristics, as compared with the bare diodes or conventional extended cavity diode lasers, that allow a significant increase in the accuracy and the sensitivity of the spectroscopic measurements. In particular, FM modulation rate of our laser is above 1~MHz,  its modulation span is above 4~GHz, and the measured RAM, at -80~dB, is at least an order of magnitude lower  than other FM modulated devices known to us.

\begin{figure}[htb]
\centerline{\includegraphics[width=8.5cm]{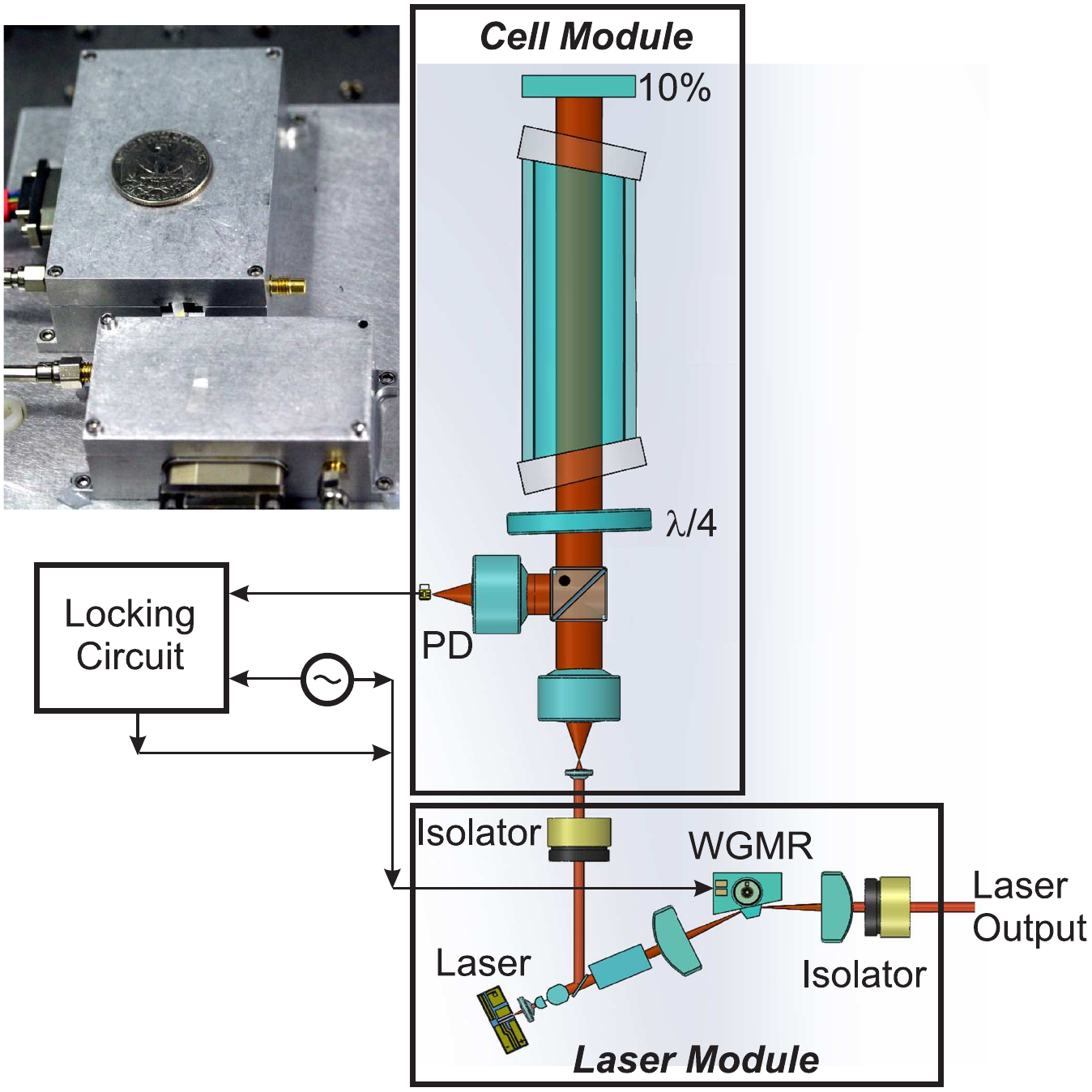}}
\caption{{\small Schematic of the experimental setup including the self-injection locked modulatable laser and a rubidium cell (the ratio between geometrical sizes of the components is conserved). Inset shows actual assembled prototype. Light from a semiconductor DFB laser (30~mW) is collimated and injected to a modulatable WGM microresonator (WGMR). The coupling efficiency is 3~dB. There is no isolator between the laser and the resonator so the laser self-injection locks to a resonator mode. Approximately 0.5~mW is split off the laser beam, expanded, and injected to the rubidium vapor cell through a polarizing beam splitter (PBS) and a quarter wave plate ($\lambda/4$). Ten percent of the light is reflected back to generate the saturated absorption resonance used to lock the device. The light exiting the cell is analyzed using photodiode PD. Signal from the photodiode is used to create an electronic feedback stabilizing the laser. The laser has 10~mW output.}\label{fig1}}
\end{figure}

To achieve these improvements, we built an external cavity distributed feedback (DFB) semiconductor laser using a high quality (Q-) factor monolithic whispering gallery mode (WGM) microresonator \cite{maleki09chap}. The modulating and dithering of the laser frequency was produced by tuning the external cavity. The frequency of the WGM resonator was modified by changing its temperature as well by stress applied via a piezo-electric transducer (PZT) actuator (see Fig.~\ref{fig1}).

We utilized theself-injection locking method for locking the laser to the WGM resonator \cite{dahmani87ol,hollberg88apl,himmerich94ao,vassiliev98oc,vassiliev03apb,liang10ol}. In this scheme the resonant stimulated Rayleigh back-scattering is used, which occurrs at the frequency of a high-Q resonator mode due to surface and volumetric inhomogeneities in the resonator \cite{weiss95ol,gorodetsky00josab}. Because self-injection locking feedback is rather fast, it results in significant reduction of the laser phase and, to a lesser extent,  amplitude noise, in a broad frequency range \cite{li88apl,li89jqe,laurent89jqe}. It also allows transferring frequency modulation from the resonator to the laser. The advantage of this transfer is that the modulation is not associated with the change of the laser power and, hence, the RAM effect is extremely small. We achieved RAM below -80~dB, which is orders of magnitude smaller than RAM with electro-optical modulators \cite{jaatinen09mst,sathian12ao}. The achieved value of RAM is in contrast to that produced by direct modulation of a DFB laser through the change of the carrier density, which unavoidably leads to a significant admixture of amplitude modulation in the phase modulated signal.

We illustrated the performance of the laser by locking it to a saturated absorption transition in rubidium (Fig.~\ref{fig1}) and demonstrated frequency stability exceeding $10^{-12}$ for integration times  from 1~s to 1~day. The achieved stability of the locked laser, combined with its compact package (50~cm$^3$) makes it useful for application in atomic clocks, magnetometers, wavelength references and other high precision applications based on frequency metrology.

\section{RAM and laser spectroscopy}

Frequency modulation and wavelength modulation laser spectroscopy \cite{moses77ol,bjorklund80ol,hall81apl,supplee94ao} benefits from usage of heterodyne methods to shift the frequency of the measured signal from the low frequency region where technical noise as well as drifts dominate,  to higher frequencies. However, residual amplitude modulation of light used in the spectroscopy mimics the signal of interest, leading to a nonzero background that limits the measurement sensitivity. In what follows we present a simplified explanation of the RAM's impact following earlier published studies \cite{supplee94ao}.

\subsection{Frequency modulation spectroscopy}

Let us consider an FM spectroscopy scheme involving a probe laser characterized with excessive intensity noise. For the sake of clarity, we consider the problem of determination of the center of a resonance having less than 100\% contrast. The complex amplitude transfer function of such a resonance can be described by the expression
\begin{equation}
F(\omega) = \frac{E_{out}}{E_{in}} = \frac{\gamma_1-\gamma_2+i(\omega_0-\omega)}{\gamma_1+\gamma_2+i(\omega_0-\omega)},
\end{equation}
where $\omega$ is the spectral frequency, $\omega_0$ is the frequency of the resonance, $E_{in}$ and $E_{out}$ are the complex amplitudes of the electric field of the input and output light waves, $\gamma_1$ and $\gamma_2$ are the parameters determining contrast, $1 \geq C \geq 0$, defined as
\begin{equation}
C=1-\frac{|E_{out}|^2}{|E_{in}|^2} \bigg \vert_{\omega=\omega_0}=\frac{4\gamma_1\gamma_2}{(\gamma_1+\gamma_2)^2},
\end{equation}
and bandwidth
\begin{equation}
\Gamma=\gamma_1+\gamma_2
\end{equation}
of the resonance.

The input light is modulated to generate an error signal. We assume that both the amplitude and phase modulation are present, and that the modulation is small
\begin{eqnarray}
E_{in}=E_0 \left (1+a \cos \Omega t \right )e^{-i(\omega t+b \sin \Omega t)} \approx \\ \nonumber
E_0e^{-i\omega t} \left [  1+ \left ( \frac a 2-\frac b 2  \right ) e^{i\Omega t} + \left ( \frac a 2+\frac b 2  \right ) e^{-i\Omega t} \right ]. \label{ein1}
\end{eqnarray}
For the amplitude of light that interacted with the resonance we find
\begin{eqnarray}
E_{out}=
E_0e^{-i\omega t} \left [  F(\omega) + \left ( \frac a 2-\frac b 2  \right ) F(\omega-\Omega) e^{i\Omega t} + \right . \\ \nonumber \left . \left ( \frac a 2+\frac b 2  \right )  F(\omega+\Omega) e^{-i\Omega t} \right ].
\end{eqnarray}
Let us assume now that the modulated light is sent to a photodiode with resistance $\rho$ and responsivity $R$, so the photocurrent $i_{PD}$ is equal to
\begin{equation}
i_{PD}=R P_{out},
\end{equation}
where $ P_{out}$ is the optical power at the photodiode (PD).
We are interested in heterodyne detection of the first modulation harmonic of the exiting RF signal
\begin{eqnarray}
\frac{P_{out}}{P_0}\bigg | _{\exp(\pm i \Omega t)}= \frac 1 2 \left [(a-b)  F(\omega) F^*(\omega-\Omega)+ \right . \\ \nonumber \left . (a+b)  F^*(\omega)  F(\omega+\Omega)  \right ]e^{-i\Omega t}+c.c.
\end{eqnarray}
To measure it, the photocurrent is electronically mixed with a local oscillator signal
\begin{equation}
i_{LO}\sim   e^{i(\Omega t +\phi_{LO} )} + e^{-i(\Omega t +\phi_{LO} )},
\end{equation}
and filtered out with a low pass filter of bandwidth $\Delta F$, to produce a DC error signal with power
\begin{eqnarray} \label{perror}
P_{error}= \rho R^2 P_0^2 \frac \mu 4 \left \{ \left [(a-b)  F(\omega) F^*(\omega-\Omega)+ \right . \right .\\ \nonumber \left . \left . (a+b)  F^*(\omega)  F(\omega+\Omega)  \right ]e^{i \phi_{LO}}+c.c. \right \}^2,
\end{eqnarray}
where $\mu$ is the mixer efficiency.

We assume that the modulation frequency is small compared with the spectral width of the resonance, $\Gamma \gg \Omega$. In this case the expression in square brackets in Eq.~(\ref{perror}) is real and, hence, we have to select $\phi_{LO}=0$. The error signal becomes
\begin{eqnarray} \label{perror1}
P_{error}= \rho R^2 P_0^2 \frac \mu 4 \left \{ 4 a \frac{(\gamma_1-\gamma_2)^2+(\omega_0-\omega)^2}{(\gamma_1+\gamma_2)^2+(\omega_0-\omega)^2} - \right . \\ \nonumber \left .    16 b \frac{\gamma_1 \gamma_2 (\omega_0-\omega) \Omega }{((\gamma_1+\gamma_2)^2+(\omega_0-\omega)^2)^2} \right \}^2,
\end{eqnarray}

We obtain the first important result of the calculation from Eq.~(\ref{perror1}): the amplitude modulation shifts the zero of the error signal by $\delta_{AM}$, that can be written in the form
\begin{equation} \label{am1}
\frac{\delta_{AM}}{\Gamma}=\frac{a}{b} \frac{1-C}{C} \frac{\Gamma}{\Omega}.
\end{equation}
It means that even small admixture of  amplitude modulation results in a significant shift of the zero of the error signal if i) $\Omega$ is much smaller than $\Gamma$ and, ii), the contrast is small, $C < 1$.

The second important result is that the error signal is given by
\begin{equation}
P_{error}= 4 \mu \rho R^2 P_0^2  b^2 C^2 \left ( \frac{\omega_0-\omega}{\Gamma}  \right )^2 \left ( \frac{\Omega}{\Gamma}  \right )^2.
\end{equation}
If we note that the power of the first modulation sideband is approximately $P_1=P_0 b^2/4$, we get
\begin{equation}
P_{error}= 16 \mu \rho R^2 P_0 P_1 C^2 \left ( \frac{\omega_0-\omega}{\Gamma}  \right )^2 \left ( \frac{\Omega}{\Gamma}  \right )^2.
\end{equation}
Let us assume that the major noise that limits the sensitivity is the relative intensity noise (RIN) at the modulation frequency. The power of the RIN-originated RF noise is
\begin{equation}
P_{RIN}= \mu \rho R^2 P_0^2 (1-C)^2 RIN \Delta F.
\end{equation}
We represent the signal to noise ratio (SNR) as $P_{error}/P_{RIN}$, and assuming that the accuracy of the lock is given by $SNR=1$, find
\begin{eqnarray} \label{det1}
\frac{|\omega_0-\omega|}{\Gamma} \approx \sqrt{\frac{P_0}{P_1}} \frac{1-C}{4C}  \frac{\Gamma}{\Omega} \sqrt{RIN(\Omega) \Delta F}.
\end{eqnarray}

\subsection{Wavelength modulation spectroscopy}

Similar result can be obtained for the case of wavelength modulation spectroscopy for which frequency deviation exceeds the modulation frequency. The phase term of the frequency modulated light can be presented in this case as
\begin{equation}
\exp -i \int \limits_0^t (\omega + \Omega_m \cos \omega_m \tau) d \tau,
\end{equation}
where $\Omega_m$ is the frequency deviation and $\omega_m\ll \Omega_m$ is the modulation frequency. We cannot use the simple harmonic decomposition utilized in Eq.~(\ref{ein1}) and instead use the following expression
\begin{eqnarray} \label{ein2}
E_{in}=E_0 \left (1+a \cos \omega_m t \right )\exp  \left [-i \left (\omega t+ \frac{\Omega_m}{\omega_m} \sin \omega_m t \right ) \right ].
\end{eqnarray}
The error signal is given in this case by the derivative of the absorption profile of the line \cite{supplee94ao}
\begin{eqnarray}
P_{error} \simeq \rho R^2 P_0^2 \mu \left ( 2 a + \frac{d |F(\omega)|^2}{d\omega} \Omega_m \right )^2 \simeq \\ \nonumber
\rho R^2 P_0^2 \mu \left [2a+ 8 \frac{\gamma_1 \gamma_2 (\omega_0-\omega) \Omega_m }{((\gamma_1+\gamma_2)^2+(\omega_0-\omega)^2)^2} \right ]^2.
\end{eqnarray}
The main approximation here is that the frequency deviation is much less than the bandwidth of the spectral line under study $\Gamma \gg \Omega_m$. The error signal decreases when $\Omega_m$ exceeds $\Gamma$ \cite{supplee94ao}.

Therefore, the center of the error signal shifts due to AM modulation as
\begin{equation} \label{am2}
\frac{\delta_{AM}}{\Gamma}=a \frac{1-C}{C} \frac{\Gamma}{\Omega_m},
\end{equation}
and the SNR limited measurement accuracy of the center of the resonance is
\begin{eqnarray} \label{det2}
\frac{|\omega_0-\omega|}{\Gamma} \approx \frac{1-C}{2C} \frac{\Gamma}{\Omega_m} \sqrt{RIN(\Omega) \Delta F}.
\end{eqnarray}
Comparing Eq.~(\ref{am1}) and Eq.~(\ref{det1}) as well as Eq.~(\ref{am2}) and Eq.~(\ref{det2}) we conclude that   if $\sqrt{RIN(\Omega) \Delta F} \gg a$ the AM modulation term does not influence the measurement accuracy, within the frame of the approximations made, . This is not the case for the majority of diode laser spectroscopy experiments. Therefore, the amplitude modulation term has to be reduced as much as possible to increase the measurement accuracy.

\section{Experiment}

In our experiment we created two identical self-injection locked lasers using WGM resonators having a diameter of 2~mm and loaded bandwidth of 500~kHz (Fig.~\ref{fig1}). The  resonator was modulated using a  PZT actuator. The PZT stresses the microresonator modifying its diameter as well as creating a change in the refraction index at the location of the optical mode through the elasto-optic effect. These effects cause the frequency of the WGMs to change. The laser tracks the changing frequency of the optical mode when self-injection locked, creating high speed frequency modulation in the optical output with very low residual amplitude noise.  The measured modulation rate was 10~MHz$/$V, with respect to the voltage at the PZT element.

\begin{figure}[htb]
\centerline{\includegraphics[width=8.5cm]{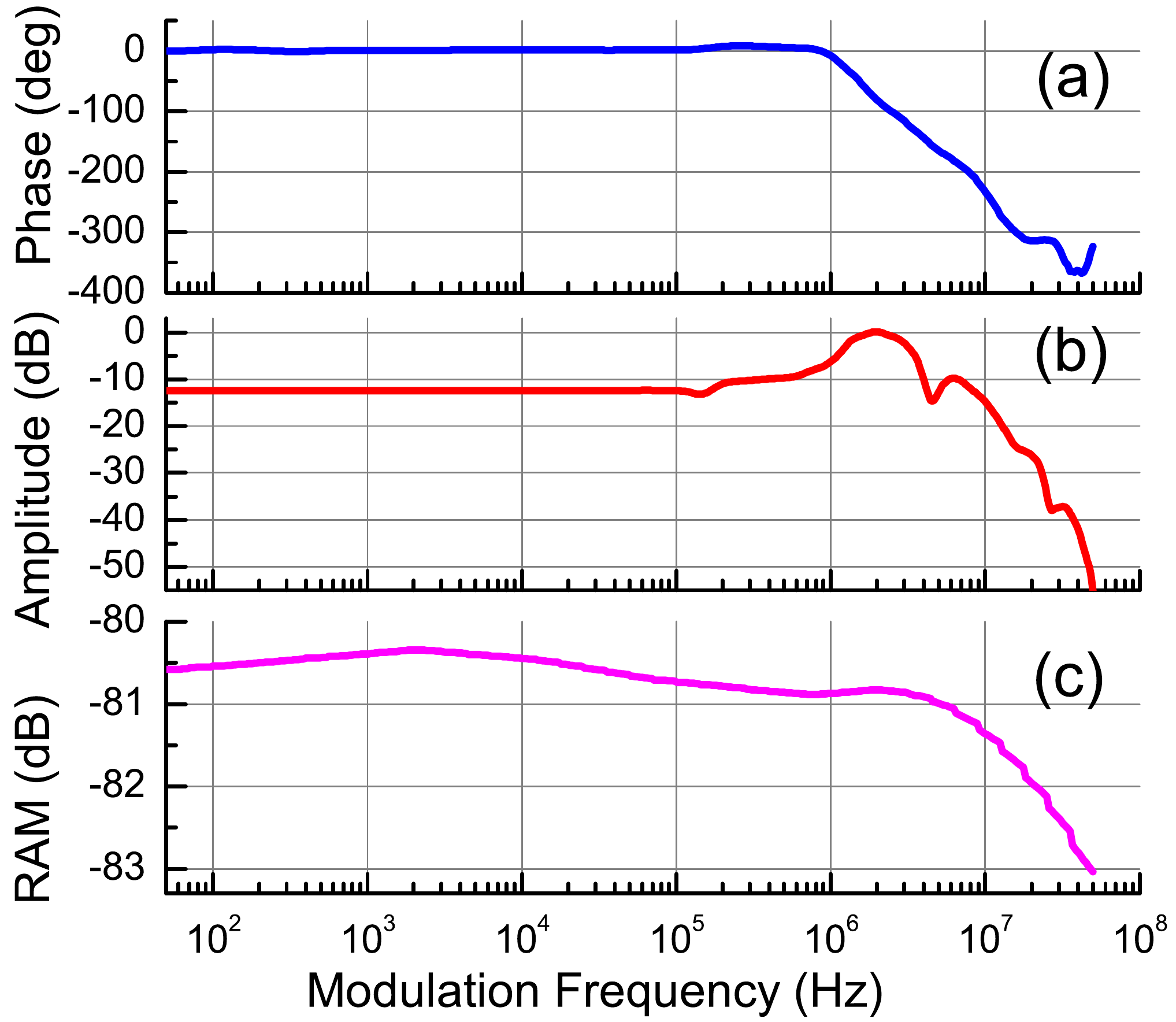}}
\caption{{\small Modulation properties of the laser. }\label{fig2}}
\end{figure}

To measure the FM response of the laser, we coupled the output light into a single mode fiber and send it through a frequency discriminator made of a Mach-Zehnder Interferometer (MZI), which converts frequency modulation to intensity modulation. A network analyzer (3577A) was used to measure the modulation frequency response. To detect the RAM associated with the FM signal we sent the light to a Thorlabs photodiode and performed the measurement using the same network analyzer. The phase and amplitude of the FM response, and the amplitude of the RAM are shown in Fig.~(\ref{fig2}). It worth noting that there are several spurs in the PZT response above 100~kHz, and the data in Fig.~(\ref{fig2}) does not have enough resolution bandwidth to see them.

\begin{figure}[htb]
\centerline{\includegraphics[width=8.cm]{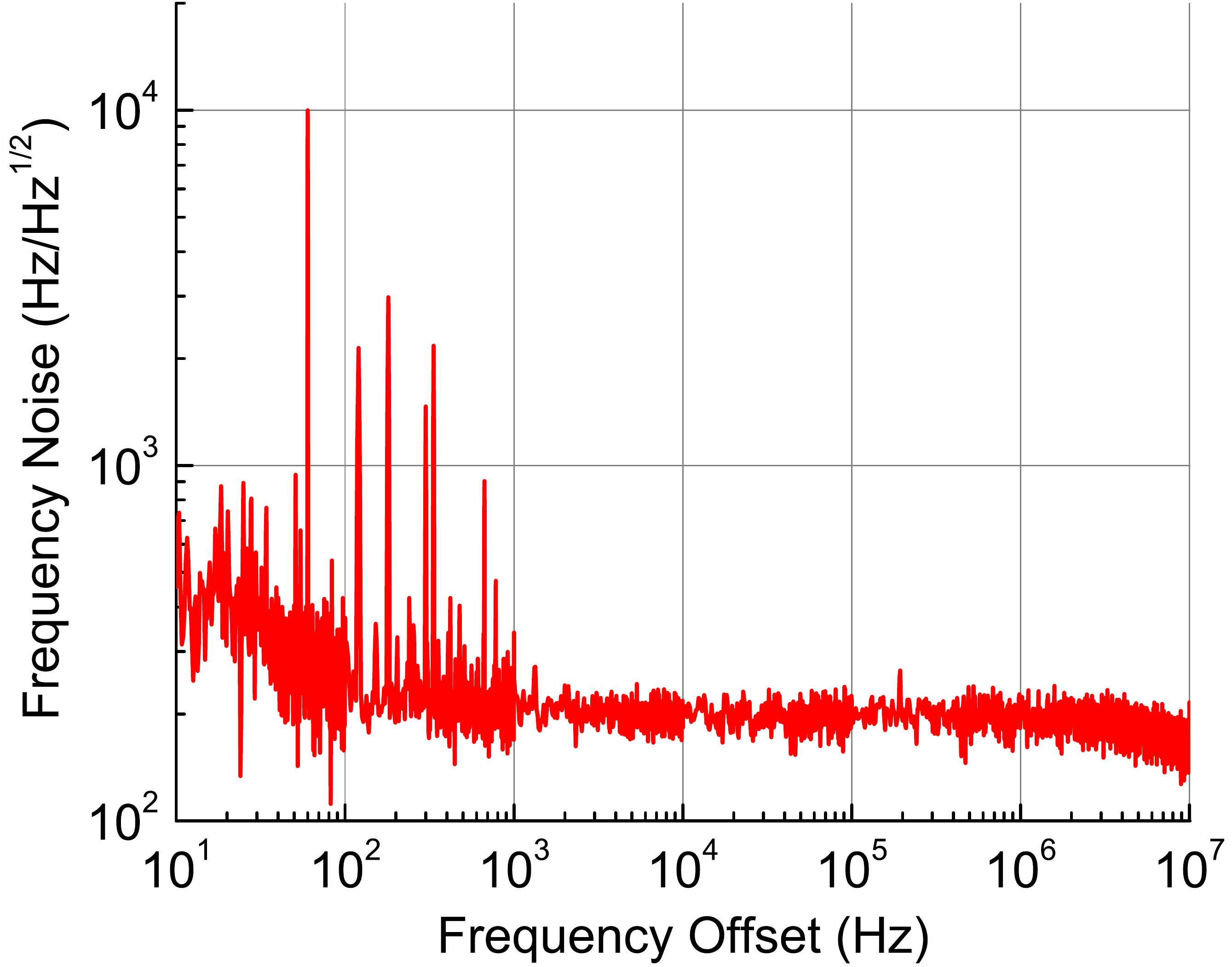}}
\caption{{\small Frequency noise of a free running self-injection locked laser. }\label{fig3}}
\end{figure}

\begin{figure}[htb]
\centerline{\includegraphics[width=8.cm]{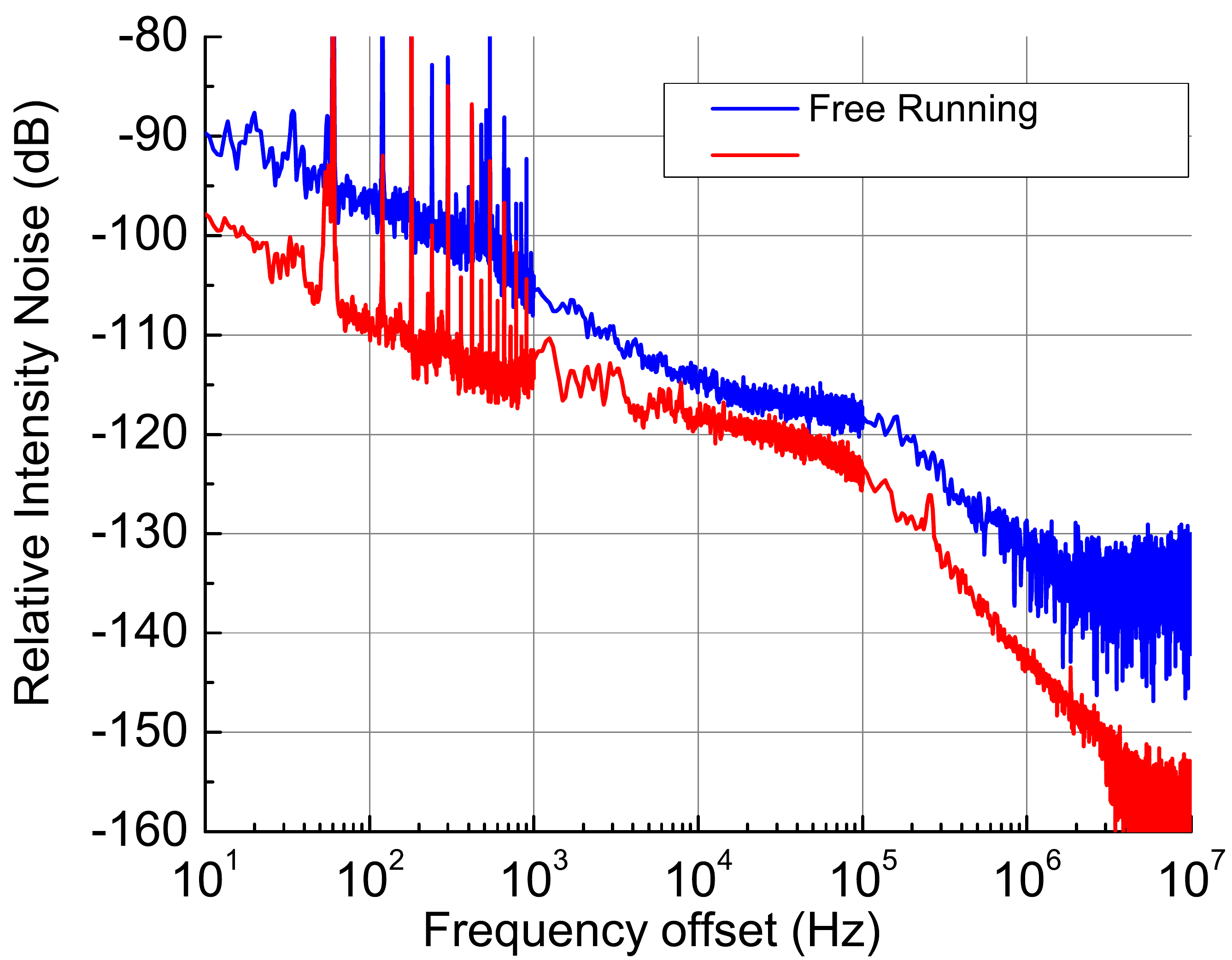}}
\caption{{\small Comparison of RIN of free running DFB laser and self-injection locked DFB laser. The noise decreases in the self-injection locked case. }\label{fig4}}
\end{figure}

To further characterize the performance of the lasers we measured their frequency noise, Fig.~(\ref{fig3}), as well as RIN,
Fig.~(\ref{fig4}). We introduced the output of both lasers toa fast photodiode and prroduced a beat signal. The beat-note radio frequency (RF) signal was centered around 10~GHz and its phase noise was measured using the RF phase noise test system (PNTS) of OEwaves. With the optical power on the photodiode at 3~mW, and the photodiode responsivity of 0.3~A$/$W, the output RF power was -17~dBm. The phase noise was converted into frequency noise and shown in Fig.~(\ref{fig3}).

To measurethe laser RIN we sent the light of one laser to a fast photodiode and measured the signal using a signal analyzer. We found that self-injection locking results in improvement of the RIN if the laser is resonantly tuned to the corresponding WGM. The RIN can increase if the laser is locked to the slope of the mode. This effect results from the FM-to-AM conversion via the resonator which is effectively a frequency discriminator.


\subsection{Locking FM lasers to rubidium transitions}

To demonstrate the suitability of the laser for FM spectroscopy, we locked the lasers to neighboring saturation absorption resonances of $D_1$ line og Rb and measured Allan deviation of the beat note of the lasers (see Fig.~\ref{fig1}). We used two independent pure $^{87}$Rb isotope cells (Triad Technology), with dimensions of 25 mm long and 10 mm diameter. The cells were kept at 38~C$^o$ using a slow surface heaters (Minco). They were also shielded against the ambient magnetic fields using a single layer $\mu$-metal sheet.

The cell windows were wedged and anti-reflection coated to reduce back-reflection. The laser light first was sent through a polarizing beam splitter (PBS) and only $\pi$-polarized light was transmitted. It then passed a quarter-wave plate before entering the cell. The light was partially reflected (10\%) at the back of the cell and went back along the same path as the incoming light. The light exiting the cell passed through the quarter-wave plate again and became $\sigma$-polarized. It was then reflected by the PBS and focused on a photodiode, which detected the saturated absorption signal needed for locking of the laser.

The laser beam in the cell had 2.8~mm in diameter and the  interrogation power was 0.5~mW. At temperature of 38~C$^o$,the resultant saturated absorbtion resonances had 20\% contrast. The lasers were frequency modulated at 100~kHz with modulation span of 2~MHz and locked to the atomic transition using the Pound-Drever-Hall technique \cite{drever83apb}. The error signals was generated from a SRS-830 lock-in amplifier and processed by a SRS PID controller. It was then combined with the modulation signal via a bias-T and fed back to the WGM PZT modulation input.
\begin{figure}[htb]
\centerline{\includegraphics[width=8.cm]{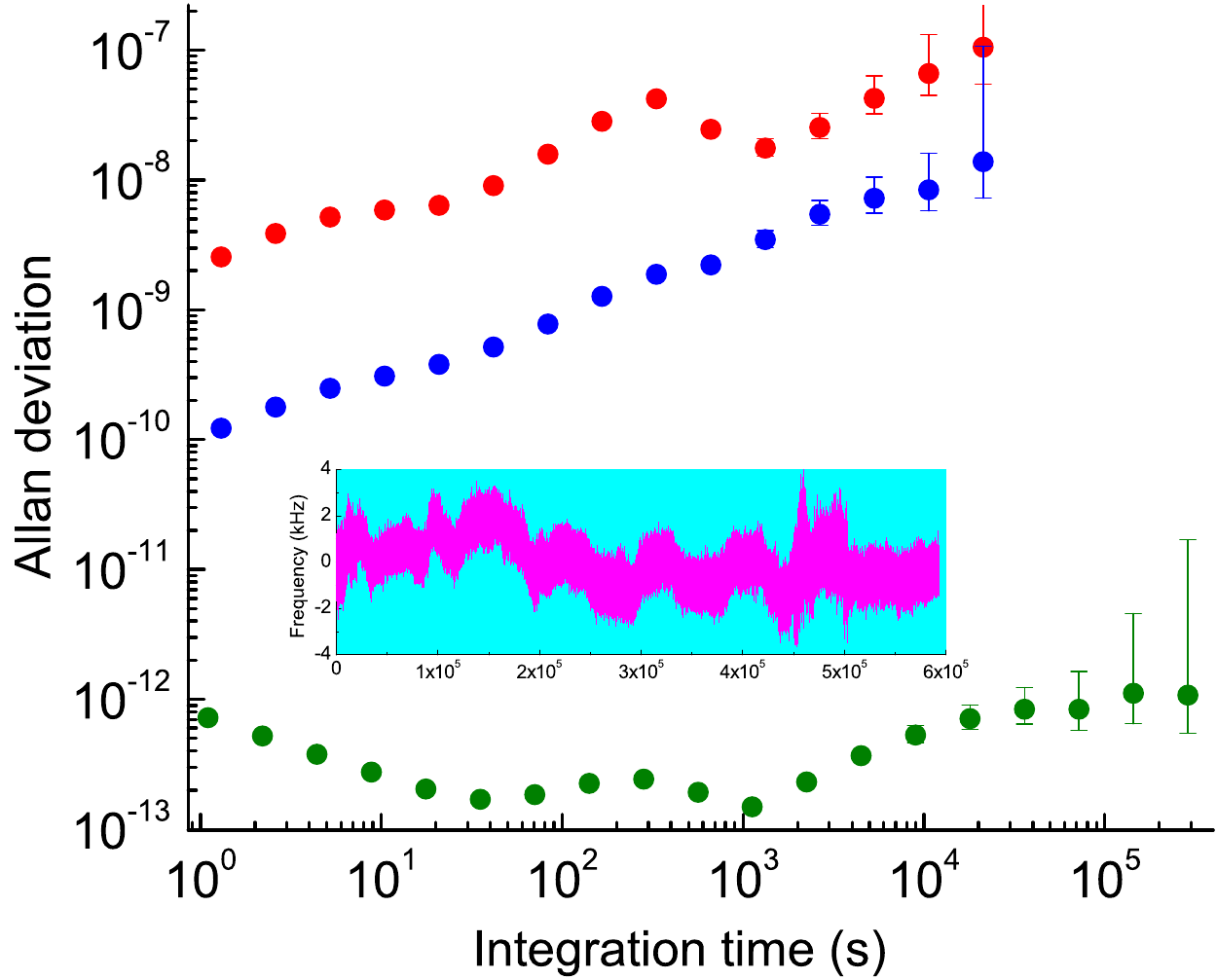}}
\caption{{\small Allan deviation of the free running DFB laser (red dots), self-injection locked DFB laser (blue dots), and self-injection locked DFB laser locked to the Doppler free rubidium transition separated by 812~MHz (green dots). The inset shows measurement trace of the frequency of the lasers' beatnote versus time that is characterized with Allan deviation shown by green dots. }\label{fig5}}
\end{figure}

The emission from the lasers was sent to a fast photodiode that generated an RF signal, stability of which was studied using a frequency counter slaved to a commercial rubidium clock. The result of the measurement is shown at Fig.~(\ref{fig5}) by green dots. The stability of the lasers' was better than $10^{-12}$ within the measurement interval of the expriment, as shown. We also repeated the measurement for the free running DFB lasers as well as the DFB lasers self-injection locked to WGM resonators. The corresponding Alan deviation is illustrated by red and blue dots in  Fig.~(\ref{fig5}). Therefore, the self-injection locking along with the electronic locking to the saturated absorption resonances of rubidium allowed improving the stability of the lasers by more than four orders of magnitude.

The stability reported in this work is higher than the stability observed with a previously demonstrated compact extended cavity diode laser locked to D$_2$ line of atomic $^{87}$Rb \cite{affolderbach05rsi,affolderbach05ole}. It is also comparable with the stability achieved in a laboratory scale diode laser setup involving Rb saturated absorption spectroscopy \cite{ye96ol}. Interestingly, the excellent stability of our lasers was achieved without fast active stabilization of the cells' temperature, which was varying by approximately 0.3~C$^o$ due to air-conditioner cycle having $\approx$1000~s period.

\section{Conclusion}
We have demonstrated a frequency modulated diode laser with reduced relative amplitude modulation. The laser allows accurate locking to rubidium saturated absorption line. We demonstrated stability of the locking exceeding $10^{-12}$ at time intervals from one second to a day. Since the modulation technology is rather generic, it can be applied to any laser self-injection locked to a WGM resonator, expanding usability of the lasers for accurate FM spectroscopy.

\section*{Acknowledgment}
Andrey Matsko acknowledges illuminating discussions with Dr. Christoph Affolderbach, Dr. Irina Novokova, and Dr. Aleksandr Zibrov.


\begin{thebibliography}{99}

\bibitem{moses77ol} E. I. Moses and C. L. Tang, "High sensitivity laser wavelength-modulation spectroscopy," Opt. Lett. {\bf 1}, 115-117 (1977).

\bibitem{bjorklund80ol} G. C. Bjorklund, "Frequency-modulation spectroscopy: a new method for measuring weak absorptions and dispersions," Opt. Lett. {\bf 5}, 15-17 (1980).

\bibitem{hall81apl} J. L. Hall, L. Hollberg, T. Baer, and H. G. Robinson, "Optical heterodyne saturation spectroscopy," Appl. Phys. Lett. {\bf 39}, 680-682 (1981).

\bibitem{whittaker85josab} E. A. Whittaker, M. Gehrtz, and G. C. Bjorklund, "Residual amplitude modulation in laser electro-optic phase modulation," J. Opt. Soc. Am. B {\bf 2}, 1320-1326 (1985).

\bibitem{gehrtz85josab} M. Gehrtz, G. C. Bjorklund, and E. A. Whittaker, “Quantum limited laser frequency-modulation spectroscopy,” J. Opt. Soc. Am. B {\bf 2}, 1510–1526 (1985).

\bibitem{wong85josab} N. C. Wong and J. L. Hall, “Servo control of amplitude modulation in frequency-modulation spectroscopy: demonstration of shot-noise-limited detection,” J. Opt. Soc. Am. B {\bf 2}, 1527–1533 (1985).

\bibitem{moerner89prl} W. E. Moerner and L. Kador, "Optical detection and spectroscopy of single molecules in a solid," Phys. Rev. Lett. {\bf 62},  2535-2538 (1989).

\bibitem{ishibashi02qels} C. Ishibashi, J. Ye, and J. L. Hall, “Analysis/reduction of residual amplitude modulation in phase/frequency modulation by an EOM,” in Conference on Quantum Electronics and Laser Science (QELS) Technical Digest Series (Institute of Electrical and Electronics Engineers, 2002), pp. 91–92.

\bibitem{duburck03tim} F. du Burck, O. Lopez, and A. El Basri, “Narrow-band correction of the residual amplitude modulation in frequency modulation spectroscopy,” IEEE Trans. Instrum. Meas. 52, 288–291 (2003).

\bibitem{duburck04mst} F. du Burck and O. Lopez, “Correction of the distortion in frequency modulation spectroscopy,” Meas. Sci. Technol. {\bf 15}, 1327–1336 (2004).

\bibitem{duburck05el} F. du Burack, A. Tabet, and O. Lopez, “Frequency-modulated laser beam with highly efficient intensity stabilisation,” Electron. Lett. 41, 188–190 (2005).

\bibitem{jaatinen09mst} E. Jaatinen, D. J. Hopper, and J. Back, “Residual amplitude modulation mechanisms in modulation transfer spectroscopy that uses electro-optic modulators,” Meas. Sci. Technol. {\bf 20}, 025302 (2009).

\bibitem{sathian12ao} J. Sathian and E. Jaatinen, "Intensity dependent residual amplitude modulation in electro-optic phase modulators," Appl. Opt. {\bf 51}, 3684-3691 (2012).

\bibitem{supplee94ao} J. M. Supplee, E. A. Whittaker, and W. Lenth, "Theoretical description of frequency modulation and wavelength modulation spectroscopy," Appl. Opt. {\bf 33}, 6294-6302 (1994).

\bibitem{hinkley71s} E. D. Hinkley and P. L. Kelly, "Detection of air pollutants with tunable diode lasers," Science {\bf 171}, 635-639 (1971).

\bibitem{bjorklund81pra} G. C. Bjorklund and M. D. Levenson, "Sub-Doppler frequencymodulation spectroscopy of I$_2$," Phys. Rev. A {\bf 24}, 166-169 (1981).

\bibitem{pokrowsky83oc} P. Pokrowsky, W. Zapka, F. Chu, and G. C. Bjorklund, "High frequency wavelength modulation spectroscopy with diode lasers," Opt. Commun. {\bf 44}, 175-179 (1983).

\bibitem{carlisle89ao} C. B. Carlisle, D. E. Cooper, and H. Preier, "Quantum noise-limited FM spectroscopy with a lead-salt diode laser," Appl. Opt. {\bf 28}, 2567-2576 (1989).

\bibitem{zhu97josab} X. Zhu and D. T. Cassidy, "Modulation spectroscopy with a semiconductor diode laser by injection-current modulation," J. Opt. Soc. Am. B {\bf 14}, 1945-1950 (1997).

\bibitem{littman78ao} M. G. Littman and H. J. Metcalf, "Spectrally narrow pulsed dye laser without beam expander," Appl. Opt. {\bf 17}, 2224-2227 (1978).

\bibitem{liu81ol} K. Liu and Michael G. Littman, "Novel geometry for single-mode scanning of tunable lasers," Opt. Lett. {\bf 6}, 117-118 (1981).

\bibitem{harvey91ol} K. C. Harvey and C. J. Myatt, "External-cavity diode laser using a grazing-incidence diffraction grating," Opt. Lett. {\bf 16}, 910-912 (1991).

\bibitem{lecomte00ao} S. Lecomte, E. Fretel, G. Mileti, and P. Thomann, "Self-aligned extended-cavity diode laser stabilized by the Zeeman effect on the cesium D$_2$ line," Appl. Opt. {\bf 39}, 1426-1429 (2000).

\bibitem{ricci95oc} L. Ricci, M. Weidemuller, T. Esslinger, A. Hemmerich, C. Zimmermann, V. Vuletic, W. Konig, and T. W. Hansch, "A compact grating-stabilized diode laser system for atomic physics," Opt. Commun. {\bf 117}, 541-549 (1995).

\bibitem{hawtorn01rsi} C. J. Hawthorn, K. P. Weber, and R. E. Scholtena, "Littrow configuration tunable external cavity diode laser with fixed direction output beam," Rev. Sci. Instruments {\bf 72}, 4477-4479 (2001).

\bibitem{maleki09chap} L. Maleki, V. S. Ilchenko, A. A. Savchenkov, and A. B. Matsko, Crystalline Whispering Gallery Mode Resonators in Optics and Photonics, Chapter 3 in “Practical Applications of Microresonators in Optics and Photonics” edited by A. B. Matsko, (CRC Press, 2009).

\bibitem{dahmani87ol} B. Dahmani, L. Hollberg, and R. Drullinger, "Frequency stabilization of semiconductor lasers by resonant optical feedback," Opt. Lett. {\bf 12}, 876-878 (1987).

\bibitem{hollberg88apl} L. Hollberg and M. Ohtsu, "Modulatable narrow-linewidth semiconductor lasers," Appl. Phys. Lett. {\bf 53}, 944-946 (1988).

\bibitem{himmerich94ao} A. Hemmerich, C. Zimmermann, and T. W. Ha\"nsch, "Compact source of coherent blue light," Appl. Opt. {\bf 33}, 988-991 (1994).

\bibitem{vassiliev98oc} V. V. Vassiliev, V. L. Velichansky, V. S. Ilchenko, M. L. Gorodetsky, L. Hollberg, and A. V. Yarovitsky, "Narrow-line-width diode laser with a high-Q microsphere  resonator", Opt. Comm. {\bf 158}, 305-312 (1998).

\bibitem{vassiliev03apb} V. V. Vassiliev, S. M. Ilina, and V. L. Velichansky, "Diode laser coupled to a high-Q microcavity via a GRIN lens", Appl. Phys. B {\bf 76}, 521-523 (2003).

\bibitem{liang10ol} W. Liang, V. S. Ilchenko, A. A. Savchenkov, A. B. Matsko, D. Seidel, and L. Maleki, "Whispering-gallery-mode-resonator-based ultranarrow linewidth external-cavity semiconductor laser," Opt. Lett. {\bf 35}, 2822-2824 (2010)

\bibitem{weiss95ol} D. S. Weiss, V. Sandoghdar, J. Hare, V. Lefevre-Seguin, J.-M. Raimond, and S. Haroche, "Splitting of high-Q Mie modes induced by light backscattering in silica microspheres," Opt. Lett. {\bf 20}, 1835-1837 (1995).

\bibitem{gorodetsky00josab} M. L. Gorodetsky, A. D. Pryamikov, and V. S. Ilchenko, "Rayleigh scattering in high-Q microspheres," J. Opt. Soc. Am. B {\bf 17}, 1051-1057 (2000).

\bibitem{li88apl} H. Li and N. B. Abraham, "Power spectrum of frequency noise of semiconductor lasers with optical feedback from a high-finesse resonator," Appl. Phys. Lett. {\bf 53},   2257-2259 (1988).

\bibitem{li89jqe} H. Li and N. B. Abraham, "Analysis of the noise spectra of a laser diode with optical feedback from a high-finesse resonator," IEEE J. Quantum Electron. {\bf 25},  1782-1793 (1989).

\bibitem{laurent89jqe} Ph. Laurent, A. Clairon, and Ch. Breant, "Frequency noise analysis of optically self-locked diode lasers," IEEE J. Quantum Electron. {\bf 25}, 1131-1142 (1989).

\bibitem{black01ajp} E. D. Black, "An introduction to Pound-Drever-Hall laser frequency stabilization," Am. J. Phys. {\bf 69}, 79-87 (2001).

\bibitem{drever83apb} R. W. P. Drever, J. L. Hall, F. V. Kowalski, J. Hough, G. M. Ford, A. J. Munley, and H. Ward. "Laser phase and frequency stabilization using an optical resonator." Applied Physics B {\bf 31}, 97-105 (1983).

\bibitem{affolderbach05rsi} C. Affolderbach and G. Mileti, "A compact laser head with high-frequency stability for Rb atomic clocks and optical instrumentation," Rev. Sci. Instruments {\bf 76}, 073108 1-5 (2005).

\bibitem{affolderbach05ole} C. Affolderbach and G. Mileti, "Tuneable, stabilized diode lasers for compact atomic frequency standards and precision wavelength references," Optics and Lasers in Engineering {\bf 43}, 291-302 (2005).

\bibitem{ye96ol} J. Ye, S. Swartz, P. Jungner, and J. L. Hall, "Hyperfine structure and absolute frequency of the $^{87}$Rb 5P$_{3/2}$ state," Opt. Lett. {\bf 21}, 1280-1282 (1996).

\end{thebibliography}
\end{document}